\documentclass{PoS}
\usepackage{cite}

\title{
\vspace*{-2.2cm}
\begin{minipage}{\textwidth}
{\normalfont\small LTH 862, DESY 10-007,~ {\tt arXiv:1001.3554 [hep-ph]}
\hspace{\fill} January 2010}\\
\end{minipage}\\[35pt]
        Higher-order predictions for splitting functions and coefficient 
        functions from physical evolution kernels$\!\!$}

\ShortTitle{Higher-order predictions from physical evolution kernels}

\author{\speaker{A. Vogt}%
        \\
        Department of Mathematical Sciences, University of Liverpool, UK\\
        E-mail: \email{Andreas.Vogt@liv.ac.uk}}

\author{S. Moch\\
        Deutsches Elektronensynchrotron DESY, Zeuthen, Germany\\
        E-mail: \email{sven-olaf.moch@desy.de}}

\author{G. Soar\\
        Department of Mathematical Sciences, University of Liverpool, UK\\
        E-mail: \email{G.N.Soar@liv.ac.uk}}

\author{J.A.M. Vermaseren\\
        NIKHEF Theory Group, Amsterdam, The Netherlands\\
        E-mail: \email{t68@nikhef.nl}}

\abstract{
\vspace*{1cm}
We have studied the physical evolution kernels for nine non-singlet 
observables in deep-inelastic scattering (DIS), semi-inclusive $e^+e^-$
annihilation and the Drell-Yan (DY) process, and for the flavour-singlet case 
of the photon- and heavy-top Higgs-exchange structure functions 
($F_2$, $F_\phi$) in DIS.
All known contributions to these kernels show an only single-logarithmic
large-$x$ enhancement at all powers of $\x1$. Conjecturing that this behaviour
persists to (all) higher orders, we have predicted the highest three (DY: two) 
double logarithms of the higher-order non-singlet coefficient functions and of
the four-loop singlet splitting functions. The coefficient-function predictions
can be written as exponentiations of $1/N$-suppressed contributions in 
Mellin-$N$ space which, however, are less predictive than the well-known 
exponentiation of the $\ln^{\,k} N$ terms.
}          

\FullConference{RADCOR 2009 - 9th International Symposium on Radiative Corrections (Applications of Quantum Field Theory to Phenomenology) \\
                 October 25-30 2009, Ascona, Switzerland} 

\newcommand{\hspn}{{\hspace{-4mm}}}
\newcommand{\hspp}{{\hspace{15mm}}}
\newcommand{\beq}{\begin{equation}}
\newcommand{\eeq}{\end{equation}}
\newcommand{\bea}{\begin{eqnarray}}
\newcommand{\eea}{\end{eqnarray}}
\newcommand{\nn}{\nonumber}
\newcommand{\ra}{\rightarrow}

\newcommand{\MSb}{$\overline{\mbox{MS}}$}
\newcommand{\as}{\alpha_{\sf s}}
\newcommand{\ar}{a_{\sf s}}
\newcommand{\ars}{a_{\sf s}^{\,2}}
\newcommand{\art}{a_{\sf s}^{\,3}}

\newcommand{\GE}{\gamma_{e}}

\def\z#1{{\zeta_{#1}^{}}}
\def\zss{{\zeta_{2}^{\,2}}}
\def\zt2{{\zeta_{2}^{\,3}}}
\def\zf2{{\zeta_{2}^{\,4}}}
\def\ca{{C_{\!A}}}

\def\cf{{C_F}}
\def\nf{{n^{}_{\! f}}}
\def\nfs{{n^{2}_{\! f}}}
\def\nft{{n^{3}_{\! f}}}
\def\cfs{{C^{\, 2}_F}}
\def\cft{{C^{\, 3}_F}}
\def\cff{{C^{\, 4}_F}}

\def\caf{{C^{}_{\!A\!F}}}
\def\cafs{{C^{\,2}_{\!A\!F}}}
\def\caft{{C^{\,3}_{\!A\!F}}}

\def\pqq(#1){p_{\rm{qq}}(#1)}
\def\pgg(#1){p_{\rm{gg}}(#1)}
\def\x1{{(1 \! - \! x)}}

\begin{document}

\section{Introduction: hard lepton-hadron processes in perturbative QCD}

\noindent
We are interested in the structure functions in deep-inelastic scattering 
(DIS), the corresponding fragmentation functions in semi-inclusive $e^+e^-$ 
annihilation (SIA), and the cross section ${1 \over \sigma_0^{}}\: d\:\! 
\sigma \!/\! d M_{ll}^{\,2}$ for Drell-Yan (DY) lepton-pair production in 
hadron-hadron collisions (see Ref.~\cite{FP82} for a detailed introduction).
These one-scale observables, here denoted by $F_a(x,Q^2)$, are generically 
given by
\beq
\label{Fa-pQCD}
 F_a(x,Q^2) \; = \;
 \big[ \, C_{a,i\{j\}}(\as(\mu^2),\mu^2\!/Q^2) \otimes f_{i}^{\,h} (\mu^2) 
 \{ \, \otimes f_{\!j}^{\,h^{\prime}} (\mu^2) \} \big] \!(x)
 + {\cal O}(1/Q^{(2)}) \:\: .
\eeq
Here $Q^2$ denotes the physical hard scale (e.g., $Q^2 = M_{ll}^{\,2}$ for the
DY case), and $x$ the corresponding scaling variable. $\mu$ represents the
\MSb\ renormalization and factorization scale (there is no need to keep them 
different here), and $\otimes$ stands for the Mellin convolution. The parts of 
Eq.~(\ref{Fa-pQCD}) in curly brackets only apply to the DY case, and summation 
over $i$ \{and $j$\} is understood. 

\vspace{1mm}
At $\mu^2 = Q^2$ the expansion of the coefficient functions $C_a$ in powers of
the strong coupling $\as$~is 
\beq
\label{Ca-exp}
 C_{a,i}(x,\as) \; = \; 
 (1 - \delta_{aL}) \,\delta_{iq}\,\delta \x1 \: + \: \ar c_{a,i}^{\,(1)}(x)
 \: + \: \ars c_{a,i}^{\,(2)}(x) \: + \: \art c_{a,i}^{\,(3)}(x) 
 \: + \: \ldots
\eeq
with $\ar \equiv \as / (4\pi)$. As indicated by the first term of the r.h.s., 
of all cases we consider here only the longitudinal coefficient functions in 
DIS and SIA vanish at order $\as^{\,0}$. The (spacelike) parton and (timelike) 
fragmentation distributions $f_{i}^{\,h}$ are, of course, non-perturbative 
quantities. However their scale dependence is calculable perturbatively via the 
renormalization-group evolution equations
\beq
\label{P-exp}
  \frac{d}{d \ln \mu^2} \: f_i^{}(x,\mu^2) \; =\; 
  [ \, P^{\,S,T}_{ik}(\as(\mu^2)) \otimes f_k^{}(\mu^2) \,](x) 
  \;\; , \quad 
  P(x,\as) \; = \; {\textstyle \sum_{\,l=0}^{}} \;\ar^{\,l+1} P^{\,(l)}(x)
  \:\: .
\eeq
Except for $F_L$ in DIS and SIA, the terms up to $c_a^{\,(n)}(x)$ and 
$P^{\,(n)}(x)$ in Eqs.~(\ref{Ca-exp}) and (\ref{P-exp}) define the N$^{\:\!n}$LO 
approximations to Eqs.~(\ref{Fa-pQCD}). Precise predictions including a sound 
numerical uncertainty estimate require, at least, calculations at the 
next-to-next-to-leading order (NNLO $\equiv$ N$^{\:\!2\:\!}$LO). The same 
order is usually required for deducing structural features such as the ones 
discussed below.

\vspace{1mm}
The NNLO coefficient functions for the quantities mentioned above Eq.~(\ref
{Fa-pQCD}), with the exception of $c_L^{\,(3)}(x)$ in SIA, have been obtained 
in Refs.~\cite{c2DIS,MVV5,c2SIA,c2DY,SMVV1,DGGL} 
(the latter two new articles deal with the only theoretically relevant 
Higgs-exchange structure function $F_\phi$ in the heavy-top limit). 
The NNLO spacelike ($S$) splitting functions in Eq.~(\ref{P-exp}) are fully 
known from Refs.~\cite{MVV34}, while for the timelike ($T$) case only the 
diagonal quantities $P^{(2)T}_{qq,gg}(x)$ have been derived so far 
\cite{P2time}.
At N$^{\:\!3\:\!}$LO only the coefficient functions for the structure functions
$F_{1,2,3,\phi}$ are available at this point \cite{MVV610,SMVV1}.

\section{\boldmath $\ln \x1$ contributions to the splitting functions and 
coefficient functions}

\noindent
From order $\as^{\,2}$ the quark coefficient functions in Eq.~(\ref{Ca-exp}) 
and quark-quark splitting functions in Eq.~(\ref{P-exp}) need to be decomposed
into (large-$x$ dominant) non-singlet and (suppressed) pure-singlet 
contributions. 
The non-singlet splitting functions receive an only single-logarithmic (SL)
higher-order enhancement, and that only in terms relatively suppressed by 
$\x1^{\, k\,\geq\, 2}$ \cite{MVV34,P2time,K89DMS05,MV5}
\beq
\label{Pns-L1}
  P_{\rm ns}^{\,(l)}(x) \; = \; A_{\:\!l+1}\, \x1_+^{-1} 
    \: + \:  B_{l+1}\, \delta \x1 \: + \: C_{l+1}\: \ln \x1
    \: + \: {\cal O} \big( \x1^{\:\!k\, \geq 1} \ln^{\:\!l} \!\x1 \big)
  \:\: .
\eeq
Also the $\cf=0$ part of the gluon-gluon splitting functions is of this form
\cite{SMVV1}. The corresponding (pure-)$\,$singlet splitting functions include 
double-logarithmic (DL) contributions 
 
\pagebreak
\vspace*{-8mm}
\beq
\label{Psg-L1}
 P_{\rm ps,gg|_\cf^{}\;}^{\,(l)} / \, P_{\rm qg,gq}^{\,(l)}\,:
 \mbox{ ~terms up to~~ }  \x1\, \ln^{\:\!2l-1} \!\x1\; /\: \ln^{\:\!2l} \!\x1
 \:\: .
\eeq

\vspace{1mm}
The non-singlet coefficient functions for the structure functions $F_{1,2,3}$, 
the (transverse, angle-integrated and asymmetric) fragmentation functions 
$F_{\:\!T\!,\,I\!,\,A}$ and the quark-antiquark annihilation DY cross section 
$F_{\:\! \rm DY}$, on the other hand, show a DL enhancement already (but not 
only) at the $\x1_+^{-1}$ plus-distribution level, i.e,
\beq
\label{Cns-L1}
 c_{a,\rm ns}^{\,(l)}\,: \mbox{ ~terms up to~ }
 \x1^{-1} \ln^{\:\!2l-1} \!\x1 \:\: .
\eeq
The highest $\x1^{-1}$ logarithms are resummed by the threshold exponentiation 
\cite{SoftGlue}, with the exponents now known to next-to-next-to-next-to-%
leading logarithmic (N$^{\:\!3\:\!}$LL) accuracy (up to the numerically 
irrelevant four-loop cusp anomalous dimension $\!A_4$ in Eq.~(\ref{Pns-L1})), 
see Refs.~\cite{MVV7,SGlueDY,SGlueSIA}. 
The leading contributions for the corresponding longitudinal DIS and SIA 
coefficient functions are down by a factor $\x1$ and one power of $\ln \x1$ 
w.r.t.~Eq.\ (\ref{Cns-L1}). Despite a recently renewed interest in such terms 
which behave as $N^{-1} \ln^{\,k} N$ in Mellin space, see, e.g., 
Refs.~\cite{OneoverN}, corresponding resummations have not been derived so far 
for these contributions to the coefficient functions.

\vspace{1mm}
In the flavour-singlet sector we will confine ourselves to the DIS cases of $F_2$ and 
$F_\phi$ \cite{SMVV1,DGGL} in the present contribution. The gluon coefficient 
function for $F_\phi$ is of the form (\ref{Cns-L1}), while the `off-diagonal' 
(see section 5) quantities are also double-logarithmic but suppressed by $\x1$,
 
\vspace{-3mm}
\beq
\label{Csg-L1} 
 c_{2,g/\phi\!,\,q}^{\,(l)}\,: \mbox{ ~terms up to~ } \ln^{\:\!2l-1} \!\x1 
 \:\: . \quad
\eeq

\section{Non-singlet physical kernels and coefficient-function predictions} 

\noindent
We now switch to moment space (and often suppress the Mellin variable $N$), 
which considerably simplifies the following calculations by turning the 
convolutions in Eqs.~(\ref{Fa-pQCD}) and (\ref{P-exp}) into simple products. 
The resulting manipulations of harmonic sums \cite{Hsum} and harmonic 
polylogarithms \cite{Hpol} have been mostly carried out in {\sc Form}$\,$3 
and {\sc TForm} \cite{Form}.

\vspace{1mm}
The non-singlet physical evolution kernels $K_a$ for the DIS and SIA cases are 
constructed by
\bea
\label{Ka-pQCD}
  \frac{d F_a}{d \ln Q^2} & \:=\: &
  \frac{d}{d \ln Q^2} \, ( C_a\, q) \; = \;
  \frac{d\, C_a}{d \ln Q^2}\, q \, + \, C_a P \, q \; = \;
  \big( \beta(\ar)\, \frac{d\, C_a}{d \ar}\, + \, C_a P \big)\, C_a^{\,-1} F_a
\nn \\[-0.5mm]
  & \:=\: &
  \Big( P_a\, + \,\beta(\ar)\, \frac{d \ln C_a}{d \ar} \Big)\, F_a
  \; = \; K_a F_a
  \; \equiv \; {\textstyle \sum_{\, l=0}^{}}\; \ar^{\, l+1} K_{a,l\,} F_a
\eea
for $\mu^2 = Q^2$ (the additional terms for $\mu^2 \neq Q^2$ can be readily
reconstructed), where $\beta(\ar)$ is the usual beta function of QCD, 
$\beta(\ar) \,=\, - \ars \beta_0 - \art \beta_1 - \ldots$ with 
$\beta_0 = 11/3\: \ca - 1/3\: \nf$ etc, and $\nf$ is the number of effectively
massless flavours. For $a \neq L$ Eq.~(\ref{Ca-exp}) leads to the expansion
\beq
\label{Ka-exp}
  K_a \;=\;
  \ar P_{a,0} \,+\: {\textstyle \sum_{\, l=1}^{}\: \ar^{\, l+1} 
  \big( P_{a,l} - \sum_{\, k=0}^{\, l-1}\; \beta_{\:\!k} \,
  \tilde{c}_{a,\, l-k} \big) }
\eeq
 
\vspace{-3mm}
\noindent
with
\vspace{-6mm}
 
\bea
\label{ctilde}
  \tilde{c}_{a,1} & = &  c_{a,1}
  \;\; , \qquad \qquad
  \tilde{c}_{a,3} \; =\; 3\, c_{a,3}^{} - 3\, c_{a,2} \, c_{a,1}
    + c_{a,1}^{\: 3}
\nn \\
  \tilde{c}_{a,2} & = &  2\, c_{a,2} - c_{a,1}^{\:2}
  \;\; , \quad
  \tilde{c}_{a,4} \; =\; 4\, c_{a,4} - 4\, c_{a,3} \, c_{a,1}
    - 2\, c_{a,2}^{\: 2} + 4\, c_{a,2} \,
   c_{a,1}^{\, 2} - c_{a,1}^{\: 4}
  \;\; , \;\; \ldots \:\: .
\eea
The structure for the DY case is the same except for $P_{a,n} \equiv 
P_{\, a, \rm ns}^{\,(n)} \,\ra\, 2\, P_{a,n}$ in Eq.~(\ref{Ka-exp}).

The threshold resummation of these coefficient functions \cite{SoftGlue}, 
again for $a \neq L$, is given by
\beq
\label{Ca-res}
  C_a(N,\as) \; = \; g_0^{}(\ar) \,\exp \:\! \{ L g_1^{}(\ar L) + g_2^{}(\ar L)
  + \ldots \} \: + \: {\cal O}(1/N)
\eeq
with $L \equiv \ln N$. Due to the logarithmic derivative in the second line of
Eq.~(\ref{Ka-pQCD}), the exponentiation (\ref{Ca-res}) guarantees a 
single-logarithmic large-$N/\,$large-$x$ enhancement of the physical kernels
\cite{NV3},
\beq
\label{Ka-res}
  K_a(N,\as) \; = \; - {\textstyle \:\sum_{\, l=1}^{} }\; A_l\, \ar^{\,l}\, L  
  \, + \, \beta(\ar) \, {d \over d \ar} \,
  \{ L g_1^{}(\ar L) + g_2^{}(\ar L) + \ldots \} \: + \: \ldots
  \:\: .
\eeq

We are now ready to present the first crucial observation: all considered 
non-singlet kernels $K_a$ (including $a = L$ in DIS and SIA) are single-%
log enhanced to all orders in $N^{-1\!}$ or $\x1$ \cite{MV3,MV5}. Switching
back to $x$-space, the universal $a \neq L$ leading-logarithmic terms in DIS 
(upper sign) and SIA (lower sign) to N$^{\:\!3\:\!}$LO read, 
with $\,\pqq(x) = \x1_+^{-1}\! -\! 1\! -\! x\,$, 
\bea
\label{Ka-LL}
 K_{a,0}^{}(x)
  & \: = \: &
      2\,\cf \pqq(x) + 3\,\cf \delta \x1
 \nn \\[0.5mm]
 K_{a,1}^{}(x)
  & \: = \: &
  \ln \x1 \,\pqq(x) \* \left[
      -2\,\cf \* \beta_0 \,\mp\, 8\,\cfs\, \ln x \right]
 \nn \\[0.5mm]
 K_{a,2}^{}(x)
  & \: = \: &
      \ln^{\,2} \!\x1 \,\pqq(x) \* \left[ \:
      2\,\cf \* \beta_0^{\,2} \,\pm\, 12\,\cfs\,\beta_0\, \ln x
      + 16\,\cft\, \ln^{\,2\!} x \right]
 \\[0.5mm]
 K_{a,3}^{}(x)
  & \: = \: &
      \ln^{\,3} \! \x1 \,\pqq(x) \* \left[
      - 2\,\cf \* \beta_0^{\,3} \,\mp\, 44/3\: \cfs\,\beta_0^{\,2}\, \ln x
      - 32\,\cft\,\beta_0\, \ln^{\,2\!} x 
      + \xi_{P_3^{}}^{}\,\cff\, \ln^{\,3\!} x
      \right]
 \nn
 \;\; , 
\eea
where $\xi_{P_3^{}}^{}$ is the unknown four-loop SL coefficient in 
Eq.~(\ref{Pns-L1}).
For DIS the N$^{\:\!3\:\!}$LO relation is based on Refs.~\cite{MVV610}, while 
for SIA we have used incomplete but sufficient analytic-continuation results 
presented in Ref.~\cite{MV5} where also the DY relations analogous to 
Eqs.~(\ref{Ka-LL}), known only to NNLO, can be found. 
The first terms on the right-hand-sides include the leading large-$\nf$ terms 
for which Eqs.~(\ref{Ka-LL}) can be generalized to all orders in DIS, using the
$C_{2,\rm ns}$ results of Ref.~\cite{Mankiewicz:1997gz}. 

\vspace{1mm}
It is now rather obvious to conjecture that the physical evolution kernels 
receive only SL contributions to all orders in $\as$ at all powers of $\nf$. 
This implies an exponentiation (see section 4) of the coefficient functions 
beyond the $\x1_+^{-1}$ terms. The emergence of the resulting fourth-order 
predictions can be illustrated by recalling the last relation written out in 
Eq.~(\ref{ctilde}),
\beq
 \underbrace{ \,\tilde{c}_{a,4}\, }_{\rm SL^{\,\!}} \:\; = \:\;
 \underbrace{ 4\, c_{a,4} }_{\rm DL,\: new^{\,\!}} \,
 \underbrace{ - \: 4\, c_{a,3} \, c_{a,1} - \,2\, c_{a,2}^{\: 2} 
   + \,4\, c_{a,2}\, c_{a,1}^{\: 2} - \, c_{a,1}^{\: 4} 
   }_{\rm DL,\: known^{\,\!}~for~DIS/SIA }
 \:\: .
\eeq
I.e., the $\ln^{\,7,6,5}\! \x1$ DL fourth-order contributions for $F_{1,2,3}$ 
and $F_{T,I,A}$ in Eq.~(\ref{Ca-exp}) need to cancel the corresponding terms 
from the known lower-order coefficient functions at all orders in $\x1$, and 
consequently can be predicted from those results. 
We do not have the space here to give an explicit example of these predictions
and their numerical size, but refer the reader to Ref.~\cite{MV5}.
Note, however, that our results explain an old observation \cite{KLS98} for the
highest $1/N$ logarithms.

\vspace{1mm}
Due to the universality of the leading terms in Eqs.~(\ref{Ka-LL}), also for
$c_{L,\rm ns}^{\,(l>3)}$ in DIS and SIA the coefficients of the three highest 
logarithms are predicted, by the respective differences $K_2 - K_1$ and 
$K_I - K_T$. 
The agreement of these predictions with those obtained from the quite different
kernels $K_L$ \cite{MV3} -- the above differences are of the order $\x1^0$, 
while the leading large-$x$ terms of $K_L$ are of the form $\x1_+^{-1}$ -- 
provides a quite non-trivial check of the above conjecture. 
On the other hand, only two logarithms can be predicted completely at this 
point at the third and all higher orders for the DY case \cite{MV5}, as the 
corresponding coefficient function is only known to order $\as^{\,2}$ 
\cite{c2DY}.

\section{All-order exponentiation of the \boldmath $1/N$ non-singlet 
coefficient functions}

\noindent
\vspace*{-1mm}
The subleading $1/N$ contributions to the non-singlet coefficient functions for
$F_{1,2,3}$, $F_{\:\!T\!,\,I\!,\,A}$ and $F_{\:\! \rm DY}$ can be cast in an
all-order form analogous to (if unavoidably less compact than)
Eq.~(\ref{Ca-res}), 
\bea
\label{Ca-1ovN}
  C_a - C_a \big|_{N^{\:\!0} L^k} & = &
  {1 \over N} \,
  \big( \big[ d_{a,1}^{\,(1)}  L + d_{a,0}^{\,(1)} \,\big] \,\ar +
  \big[ d_{a,1}^{\,(2)}\, L + d_{a,0}^{\,(2)} \big]
        \,\ar^{\,2} + \ldots \big)
  \, \exp \left\{ L\:\! h_1(\ar L) + h_2(\ar L) + \ldots^{\,} \right\}
 \; .\nn \\[-1mm] & &
\eea
 
\vspace{-2mm}
\noindent
The exponentiation functions are defined by the series $\, h_{k}(\ar L) 
\, = \, \sum_{\,k=1}^{} h_{kn} (\ar L)^n$ with $L \equiv \ln N$. 
Their coefficients for DIS$/$SIA (given by the upper$/$lower sign in Eqs.\
(\ref{ha-exp}) and (\ref{CL-res})) relative to the corresponding coefficients 
for the $N^{\:\!0} L^k$ soft-gluon exponentiation are given by 
\bea
\label{ha-exp}
 h_{1k}^{} & \:=\: & g_{1k}^{}
 \:\: , \hspace*{2.6cm}
 h_{22}^{} \;=\;  g_{22}^{} \,+\, {5 \over 24}\, \beta_0^2
  \,\pm\, {17 \over 9}\, \beta_0\,\cf \,-\, 18\,\cfs
 \\
 h_{21}^{} & \:=\: & g_{21}^{} \,+\, {1 \over 2}\, \beta_0 \,\pm\, 6\,\cf
 \:\: , \;\;\;
 h_{23}^{} \;=\;  g_{23}^{} \,+\, {1 \over 8}\, \beta_0^{\,3}
  \,\pm\, \left(\, {\xi_{\rm K_4}^{} \over 8} - {53 \over 18} \,\right)
  \beta_0^{\,2} \cf \,-\, {34 \over 3}\, \beta_0\,\cfs \,\pm\, 72\,\cft
 \:\: .\nn
\eea
Note that only the $\cf \beta_0^{\,l}$ and $\cfs\, \beta_0^{\,l-1}$ terms of
$K_{a,\,l}$ in Eqs.~(\ref{Ka-LL}) are relevant at this order in $1/N$. 
$\xi_{\rm K_4}^{}$ is the corresponding subleading large-$\nf$ coefficient at 
the fourth order, the calculation of which should become possible in the not
too distant future. Also the first term of $h_{3}^{}$ in Eq.~(\ref{Ca-1ovN}) is
known for DIS and SIA but non-universal, as the effects of $F_L$ set in at this
point. See again Ref.~\cite{MV5} for these results as well as the prefactor 
coefficients in Eq.~(\ref{Ca-1ovN}) and all corresponding DY results.

\vspace{1mm}
The corresponding exponentiation for the longitudinal structure function and
fragmentation function is given by~\cite{MV3} 
(see also Ref.~\cite{Akhoury:1998gs})
\bea
\label{CL-res}
  C_{L}^{\:\! (\pm)} (N) \: = \:
  N^{\,-1} ( d_{\:\!L,1}^{\, (\pm)} \ar + d_{\:\!L,2}^{\, (\pm)} \ar^2 
  + \ldots )
  \, \exp \left\{ L\:\! h_{L,1}(\ar L) + h_{L,2}(\ar L) + \ldots^{\,} \right\}
  \: + \: {\cal O}(N^{\,-2}) \; ,
\eea
where the following coefficients can be determined from the third-order
result of Refs.~\cite{MVV5,MV5}:
\bea 
\label{hL-exp}
  h_{\:\!L,11}^{} & \:=\: & 2\,\cf
\:\: , \quad
  h_{\:\!L,12}^{} \:\: = \:\: \frac{2}{3}\: \beta_0\, \cf
\:\: , \quad
  h_{\:\!L,13}^{} \:\: = \:\: \frac{1}{3}\: \beta_0^{\,2}\, \cf
 \nn \\
  h_{\:\!L,21}^{} & \:=\: & \beta_0 \:+\: 4\,\GE\cf \:-\: \cf
    \:+\: (4 - 4\,\zeta_2) (\ca - 2\cf)
 \nn \\
  h_{\:\!L,22}^{} & \:=\: &
  {1 \over 2}\, (\, \beta_0 \,h_{21}^{} + A_2 ) 
  \: - \:
  { 8\, (\ca - 2\cf)^2 ( 1 - 3\,\z2 + \z3_{\,\!} + \zss\, ) }
  \:\: .
\eea
 
Both these exponentiations have far less predictive power than their
$N^{\:\!0} L^k$ counterparts \cite{SoftGlue,MVV7,SGlueDY,SGlueSIA} where, e.g.,
unlike in Eqs.~(\ref{ha-exp}) and (\ref{hL-exp}), no other new coefficient 
enters $g_{22}^{}$ besides the two-loop cusp anomalous dimension $A_2$. A full
NLL accuracy, i.e., a complete determination of the function $h_{2}^{}(\ar L)$
may be feasible for Eq.~(\ref{Ca-1ovN}). On the other hand, the corresponding 
leading coefficients for $h_3^{}$ \cite{MV5} and the results for 
$h_{\:\!L,2}^{}$ in Eq.~(\ref{hL-exp}) indicate a major, possibly insurmountable
obstacle on the way to full NNLL and NLL accuracy for the quantities 
$F_a$ ($a \neq L$) and $F_L$, respectively.

\section{The singlet evolution of $F_2$ and $F_\phi$ and splitting-function 
predictions}

\vspace*{-1mm}
\noindent
DIS via the exchange of a scalar $\phi$ directly coupling only to gluons
(like the Higgs boson in the heavy-top limit \cite{HGGeff}), is an ideal 
complement to the standard structure function $F_2$. The evolution kernels for 
the resulting system of observables are as in the first line of 
Eq.~(\ref{Ka-pQCD}), 
but with 
\bea
  F \:=\:
  \Big( \begin{array}{c}  F_2 \\[-1mm] F_\phi \end{array} \Big)
\; , \;\;\;
  C \:=\:
  \Big(  \begin{array}{cc} C_{2,q}^{} & C_{2,g}^{} \\[-0.5mm]
  C_{\phi\!,\,q} & C_{\phi\!,\,g} \end{array} \Big)
\; , \;\;\;
 K \:=\:
 \Big( \begin{array}{cc} K_{22} & K_{2\phi} \\[-0.5mm]
  K_{\phi 2} & K_{\phi\phi} \end{array} \Big)
\eea
and the splitting-functions matrix $P_{\,ij}$. This system has first been 
discussed at NLO in Ref.~\cite{FP82} (it may also be interesting to study other
systems such as ($F_2$, $F_L$) \cite{F2FL} and corresponding SIA cases).
Instead of the second line of Eq.~(\ref{Ka-pQCD}), we now have (with 
$[\:\! C,P\:\! ]$ denoting the matrix commutator) 
\beq
\label{Ksg-DL}
  \frac{d F}{d \ln Q^2} \;=\;
  \big( \, \underbrace{ \beta(\ar)\, \frac{d \ln C}{d \ar}_{\,\!} }_{}
  \: + \: \underbrace{ [\:\! C,P\:\! ]_{\,\!} \, C^{\,-1} + P }_{}
  \big)\,  F \; = \; K\, F
  \:\: .
\eeq  

\vspace{-5.5mm}
\hspace*{4.6cm}{\small DL (ns + ps)\hspace{4.5mm} DL (singlet only)}

\vspace{2mm}
As far as they are completely known now, i.e., at NLO and NNLO, also the matrix
entries of $K$ show an only single-logarithmic enhancement at all powers of
$\x1$, $K_{ab}^{(n)} \,\sim\, \ln^{\,n}\! \x1 + \dots\,$. Moreover, the
leading-log contributions to $K_{22/ \phi\phi}^{(n)}$ are the same as in the
non-singlet quark-case and the very closely related $\cf=0$ gluon case 
\cite{SMVV1}.
Conjecturing that this behaviour holds also at N$^{\:\!3\:\!}$LO, the highest 
three logarithms of the unknown four-loop splitting functions,
\beq
 \ln^{6,5,4}\! \x1\, \mbox{ ~of~ } P_{\rm qg,gq}^{(3)} \mbox{ ~~~and~ }
 \ln^{5,4,3}\! \x1\, \mbox{ ~of~ } P_{\rm ps,gg|_\cf^{}}^{(3)}
\eeq
can be predicted from the known \cite{MVV610,SMVV1} three-loop coefficient 
functions for $F_2$ and $F_\phi$ at all orders in $\x1$. For example,  
the leading $\x1^0$ part of the N$^{\:\!3\:\!}$LO gluon-quark splitting 
function reads
\bea
\label{Pqg3DL0}
  P_{\rm qg}^{\,(3)}(x) &\: = \:&
     \ln^{\,6}\! \x1 \: \cdot \, 0
     \; + \; 
     \ln^{\,5}\! \x1 \* \Big[\,
            {22 \over 27}\: \* \caft \* \nf
          \,- \, {14 \over 27}\: \* \cafs \* \cf \* \nf
          \,- \, {4 \over 27}\: \* \cafs \* \nfs
          \Big]
\nn \\[1mm] & & \mbox{\hspn} \!
     + \; \ln^{\,4}\! \x1 \* \Big[\, \Big( \, {293 \over 27}
            \,-\, {80 \over 9}\: \* \z2\! \Big)\, \* \caft \* \nf
          \,+\, \Big( \, {4477 \over 162}
          \,-\, 8 \* \z2\! \Big)\, \* \cafs \* \cf \* \nf
          \,-\, {13 \over 81}\: \* \caf \* \cfs \* \nf
\nn \\[1mm] & & \mbox{\hspp}
          \,-\, {116 \over 81}\: \* \cafs \* \nfs
          \,+\, {17 \over 81}\: \* \caf \* \cf \* \nfs
          \,-\, {4 \over 81}\: \* \caf \* \nft
          \Big]
     \; + \; {\cal O} \left( \ln^3 \! \x1 \right)
\eea
with $\caf \equiv \ca - \cf$.
The vanishing of the leading $\ln^{\,6} \x1$ term is due to an accidental
cancellation of positive and negative contributions to its coefficient.
On the other hand, the colour factors of the DL terms in Eq.~(\ref{Pqg3DL0})
follow the same pattern as the corresponding lower-order contributions: all
DL terms vanish for $\ca = \cf$ (part of the supersymmetric limit), with the
leading terms of $P_{\rm qg}^{\,(l)}$ being of the form 
$\nf\, C_{\!A\!F}^{\:l}$, the next-to-leading logarithms $\nf \{\cf, \nf\}\, 
C_{\!A\!F}^{\:l-1}$ etc.
Rather non-trivially, this pattern is predicted to hold for all four-loop 
singlet splitting functions at all orders in $\x1$ \cite{SMVV1}.

\vspace{1mm}
Unlike in the non-singlet case, there is no direct all-order generalization
here, as the cancellation of the DL contributions in Eq.~(\ref{Ksg-DL}) 
involves the corresponding terms (\ref{Psg-L1}) and (\ref{Csg-L1}) of the 
N$^{\:\!n}$LO splitting functions and coefficient functions which are both 
unknown at $n \geq 4$. One may try a simultaneous extraction of at least the
leading logarithms of both quantities, but it turns out that the only 
single-logarithmic enhancement of the physical kernel does not quite provide 
enough constraints, even if the colour structure of the previous paragraph is 
assumed in addition.

\end{document}